\documentclass[smallcondensed]{svjour3}     % onecolumn (ditto)
\usepackage{graphicx}
\usepackage{amsfonts}
\begin{document}

\title{Reflection identities of harmonic sums up to weight three. 
\thanks{The paper is dedicated to the memory of Lev Lipatov.}
}


\author{Alex Prygarin   %     \and         Second Author %etc.
}

\institute{Alex Prygarin \at
              Department of Physics \\
              Ariel University\\
              Ariel, 40700, Israel \\
              Tel.: +972-3-9066270\\
              Fax: +972-3-9066395\\
              \email{alexanderp@ariel.ac.il}           %  \\
          % \and
          % S. Author \at
               %second address
}

\date{}

\def\makeheadbox{\relax}
\maketitle

\begin{abstract}

We discuss  reflections identities of  harmonic sums up to weight three. The need for this kind of identities emerges in analysis of the general structure of eigenvalue of the BFKL equation. The reflection identities decompose a product of two harmonic sums with 
pole singularities at  real integer  points  into a linear combination of other functions with pole singularities at either negative integers or zero and positive integers. This provides a pole separation of expressions with a mixed pole structure.  
\vspace{1cm} 
\\
%PACS:02.30.Gp,02.30.Lt,02.30.Mv,02.30.Sa, 12.38.Bx, 12.40.Nn, 12.60.Jv
%\\
%Insert your abstract here. Include keywords, PACS and mathematical
%subject classification numbers as needed.
%\keywords{Harmonic Sums \and Reflection Identity \and BFKL equation}
\end{abstract}

\section{Introduction}
\label{intro}
The finite harmonic sums~\cite{HS1,Vermaseren:1998uu,Blumlein:1998if,Remiddi:1999ew} are widely used in the perturbative calculations of the gauge theories in Particle Physics. The harmonic sums are rich of identities and very convenient for writing the functional basis of possible solutions because of the fact that the weight of harmonic sums corresponds to the definite transcendentality of the resulting function.  The empiric principle of maximal transcendentality in supersymmetric quantum field theories~\cite{Kotikov:2006ts} was used for guessing basis functions in calculating the eigenvalues of the Balitsky-Fadin-Kuraev-Lipatov~(BFKL) equation in $\mathcal{N}=4$ super Yang-Mills theory~(SYM). The results of calculations were presented in the terms of real functions of complex variable $F(z)$, where the full real-valued answer was written as $F(z)+F(\bar{z})$ for specific values of $z$.  All attempts of writing the full answer for a general $z$ were not successful  suggesting that one should search other form of presenting the final result. One of the options would be to use a product 
of two real functions of complex variable $f(z)g(\bar{z})$ and its linear combination. In the course of checking that option we faced a need for a special 
kind of identities not currently available for the harmonic sums namely the ones where a product of two functions  $f(z)g(-1-z)$ can be represented as a sum of two other functions of the same arguments $F(z)+G(-1-z)$. We call those identities the reflection identities because the argument is reflected with respect to point $\frac{1}{2}(z+(-1-z))=-\frac{1}{2}$. 

In this paper we mainly focus on the reflection identities for the harmonic sums at weight $w=2$ and $w=3$, leaving discussion of the physical motivation and their possible applications for     future studies.

%the Appendix~\ref{APP:BFKL}. 

The paper is structured as follows. In the first Section we discuss the definition of the harmonic sums and their  analytic continuation to the complex plane.  In the second Section we introduce the reflection identities and discuss their compatibility with the known identities. We present the list of obtained reflection identities at weight two and three.
%  Some technical aspects of the calculations are presented in the Appendices. 

\section{Harmonic Sums}
\label{sec:HS}
The harmonic sums are defined~\cite{HS1,Vermaseren:1998uu,Blumlein:1998if,Remiddi:1999ew} through nested summation for $n\in \mathbb{N}$
\begin{eqnarray}\label{defS}
S_{a_1,a_2,...,a_k}(n)=  \sum_{n \geq i_1 \geq i_2 \geq ... \geq i_k \geq 1 }   \frac{\mathtt{sign}(a_1)^{i_1}}{i_1^{|a_1|}}... \frac{\mathtt{sign}(a_k)^{i_k}}{i_k^{|a_k|}}
\end{eqnarray}

In this paper we consider the harmonic sums with only real integer values of $a_i$, which build the alphabet of the possible negative and positive indices.  
In eq.~\ref{defS} $k$ is  the depth and $w=\sum_{i=1}^{k}|a_i|$ is the weight of the harmonic sum $S_{a_1,a_2,...,a_k}(n)$. 

The indices of harmonic sums $a_1,a_2,...,a_k$ can be either positive or negative integers and label uniquely $S_{a_1,a_2,...,a_k}(n)$ for any given 
weight. However there is no unique way of building the functional basis for a given weight because the harmonic sums are subject to so called shuffle relations, where a linear combination of $S_{a_1,a_2,...,a_k}(n)$ with the same argument but all possible permutations of indices can be expressed through 
a non-linear combinations of harmonic sums at lower weight.
 There is also some freedom in choosing the irreducible minimal set of $S_{a_1,a_2,...,a_k}(n)$ that builds those non-linear combinations.  
The shuffle relations relate between the linear and non-linear combinations of the harmonic sums of the same argument. For example,  the shuffle relation at depth two is given by
\begin{eqnarray}
S_{a,b}(z)+S_{b,a}(z)= S_{a}(z) S_{b}(z)+S_{\textrm{ sign}(a) \textrm{ sign}(b)(|a|+|b|)}(z)
\end{eqnarray}
The shuffle relations of the harmonic sums is closely connected to the shuffle algebra of the  harmonic polylogarithms~\cite{Remiddi:1999ew}.

There is another type of identity called the duplication identities where a combination of harmonic sums of argument $n$ can be expressed through 
a harmonic sum of the argument $2 n$. The duplication identities introduce another freedom in choosing the functional basis.

The definitions, functional identities, general properties of the harmonic sums and their generalizations are discussed in more details in Refs.~\cite{Blumlein:1998if,AblingerThesis,Ablinger:2011te,Blumlein:2009ta,Blumlein:2009fz} and here we focus only on the so called linear basis of the harmonic sums where we take into account only $S_{a_1,a_2,...,a_k}(n)$ with all possible permutations of the indices without applying any kind of functional identities. The number of elements in such a linear  basis equals the number of elements in any other  basis built of harmonic sums where one applies either shuffle or duplication identities. Ours choice of the  linear basis is merely a matter of convenience for the purpose of the present discussion. 

The harmonic sums $S_{a_1,a_2,...,a_k}(n)$ are defined for positive integer argument $n$ and require an analytic continuation to the complex plane if one wishes to use them as a general functional basis. There are two different analytic continuations of the harmonic sums  $S_{a_1,a_2,...,a_k}(n)$ with at least one negative index $a_i$: the analytic continuation from the even integer values of the argument $n$, which we denote by $\bar{S}^{+}_{a_1,a_2,...,a_k}(n)$, and the analytic continuation from the odd integer values of $n$, which we denote by $\bar{S}^{-}_{a_1,a_2,...,a_k}(n)$.  The details of the analytic continuation from either odd or even values of $n$ to all integers are presented in Ref.~\cite{Kotikov:2005gr}. The analytic continuation to the complex plane are then  done in terms of the Mellin transform of the Harmonic Polylogartithms~(HPL) as discussed  in Ref.~\cite{Blumlein:1998if}.

%More details  on the analytic continuation are presented in the Appendix~\cite{app:cont}.

\section{Reflection Identities}
\label{sec:reflection}
The harmonic sums $S_{a_1,a_2,...,a_k}(n)$ defined in eq.~\ref{defS} after the analytic continuation have pole singularities for the negative values of the argument. As it was mentioned in the previous Section there are two possible analytic continuations to the complex plane, one from odd integer $n$ and another from even integer $n$. For $S_{a_1,a_2,...,a_k}(n)$ with at least one negative $a_i$ those two analytic continuations represent two different functions. In recent studies~\cite{Gromov:2015vua,Caron-Huot:2016tzz,Alfimov:2018cms,Bondarenko:2015tba,Bondarenko:2016tws} related to calculations of high order perturbative corrections to the BFKL eigenvalue the analytic continuation from even values of $n$ is used. The present paper presents a step towards understanding a general structure of the BFKL eigenvalue and is based on those studies, so that we adopt their notation and consider only analytic continuation from even integer $n$ throughout this paper. In further discussions we drop the "bar-plus" notation of  $\bar{S}^{+}_{a_1,a_2,...,a_k}(n)$ for even $n$ and  use  $S_{a_1,a_2,...,a_k}(n)$ for any harmonic sum after its analytic continuation from even integer $n$ to complex plane. 

The main objective of the present paper is to introduce the reflection identities not yet available in the literature. By reflection identity we mean a functional identity where a product of two harmonic sums, one of argument $z$ and another of argument $-1-z$ can be written as a linear combination of another sums of argument $z$ and $-1-z$. As an example, consider 
\begin{eqnarray}\label{s1sc1}
S_1(z)S_1(-1-z)=S_{1,1}(z)+S_{1,1}(-1-z)+\frac{\pi^2}{3}.
\end{eqnarray} 
The sum $S_1(z)$ has pole singularities  at negative integers whereas the sum $S_1(-1-z)$ has pole singularities at zero and positive integers. The product 
of two functions on the left hand side of eq.~\ref{s1sc1} have poles at all integers and zero, while each individual term at the right hand side has only pole on  either negative integers or zero and positive integers. This way we perform a pole separation of the product $S_1(z)S_1(-1-z)$ with a mixed pole structure.  We call eq.~\ref{s1sc1} the reflection identity because the argument of the harmonic sum is reflected with respect to a point 
\begin{eqnarray}
\frac{z+(-1-z)}{2}=-\frac{1}{2}
\end{eqnarray}
The reflection identity given in eq.~\ref{s1sc1} can be written in terms of the well known digamma function~($\psi(z)=\frac{\ln \Gamma(z)}{dz}$) and its derivative as follows
\begin{eqnarray}
&&
(\psi(1+z)-\psi(1))(\psi(-z)-\psi(1))= \frac{1}{2}(\psi(1+z)-\psi(1))^2-\frac{1}{2}\psi'(1+z) \nonumber
\\
&&\hspace{2cm}+\frac{1}{2}(\psi(-z)-\psi(1))^2-\frac{1}{2}\psi'(-z)+\frac{\pi^2}{2}
\end{eqnarray}

\subsection{Weight Two}\label{sec:w2}

At weight two~($w=2$) there are  only three  reflection identities that are constructed by taking products of two basis functions at weight one
\begin{eqnarray}\label{b1}
\left\{S_{-1}(z),S_1(z) \right\}
\end{eqnarray}
The reflection identities at  $w=2$ are given by
\begin{eqnarray}\label{NEWs1sc1}
S_1(z)S_1(-1-z)=S_{1,1}(z)+S_{1,1}(-1-z)+\frac{\pi^2}{3}
\end{eqnarray}

\begin{eqnarray}\label{sm1scm1}
&& S_{-1}(z)S_{-1}(-1-z)=S_{-1,-1}(z)+S_{-1,-1}(-1-z)\nonumber
\\
&& \hspace{2cm}- 2 S_{-1}(z)\ln ( 2) - 2 S_{-1}(-1-z)\ln( 2)
 +
\frac{\pi^2}{6}-2 \ln^2 ( 2)
\end{eqnarray}

\begin{eqnarray}\label{sm1sc1}
&& S_{-1}(z)S_1(-1-z)=S_{1,-1}(z)-S_{-1,1}(-1-z)\nonumber
\\
&& \hspace{2cm}-  S_{-1}(z)\ln ( 2) + S_{1}(z)\ln ( 2)-  S_{-1}(-1-z)\ln ( 2) \nonumber
\\
&& \hspace{2cm}+ S_{1}(-1-z)\ln ( 2)
 -
\frac{\pi^2}{12}- \ln^2 ( 2)
\end{eqnarray}
All other other reflection identities at $w=2$ are obtained by  a change of  the argument $z \leftrightarrow -1-z$.

The reflection identities at weight two presented here are not new and were known long time ago in the context of functions related to the Euler Gamma function. To the best of our knowledge they appear the  eraliest in  Chapter 20 of Ref.~\cite{handbuch} and then were related to the harmonic sums~(see eqs.6.11-6.15 of  Ref.~\cite{Blumlein:2009ta}).  
 
 Similar reflection identities at weight three and up were overlooked to the best of our knowledge and are first presented in this paper.  
 
\subsection{Weight Three}\label{sec:w3}
The reflection identities at weight three are constructed by corresponding products of two basis functions at weight in  eq.~\ref{b1} and six basis functions at weight two given by 
\begin{eqnarray}
\left\{S_{-2}(z),S_2(z),S_{-1,1}(z),S_{1,-1}(z),S_{1,1}(z),S_{-1,-1}(z)\right\}
\end{eqnarray}
The resulting twelve reflection identities at weight three read

 % \{S(-1) SC(-2) ,S(-1) SC(2) ,S(-1) SC(-1,1) ,S(-1) SC(1,-1) ,S(-1) SC(1,1) ,S(-1) SC(-1,-1) ,S(1) SC(-2) ,S(1) SC(2) ,S(1) SC(-1,1) ,S(1) SC(1,-1) ,S(1) SC(1,1) ,S(1) SC(-1,-1) \}

\begin{eqnarray}
 && S_{-1}(z) S_{-2}(-1-z) = -\frac{1}{6} \pi ^2 \log (2)+\frac{9 \zeta (3)}{4}
 -\log (2)
   S_{-2}(-1-z)+\log (2) S_{-2}(z) 
   \nonumber
    \\
 && 
 \hspace{1.9cm}-\frac{1}{12} \pi ^2 S_{-1}(-1-z)-\frac{1}{12} \pi ^2
   S_{-1}(z)-\log (2) S_2(-1-z)
      \nonumber
    \\
 && 
 \hspace{1.9cm}-\log (2) S_2(z)+S_{-2,-1}(z)-S_{-1,-2}(-1-z)
\end{eqnarray}

\begin{eqnarray}
&& S_{-1}(z) S_2(-1-z)= -\frac{1}{6} \pi ^2 \log (2)-\frac{3 \zeta (3)}{4}-\log (2)
   S_{-2}(-1-z)
   +\log (2) S_{-2}(z)
   \nonumber
    \\
 && 
 \hspace{1.9cm}
 -\frac{1}{12} \pi ^2 S_{-1}(-1-z)-\frac{1}{12} \pi ^2
   S_{-1}(z)-\log (2) S_2(-1-z)
   \nonumber
    \\
 && 
 \hspace{1.9cm}-\log (2) S_2(z)-S_{-1,2}(-1-z)-S_{2,-1}(z)
\end{eqnarray}

\begin{eqnarray}
&& S_{-1}(z) S_{-1,1}(-1-z)= \frac{1}{12} (-5) \pi ^2 \log (2)+\log ^3(2)+\frac{9 \zeta
   (3)}{4}+\log (2) S_{-2}(z)
   \nonumber
    \\
 && 
 \hspace{1.9cm}
 -\frac{1}{6} \pi ^2 S_{-1}(-1-z)-\frac{1}{12} \pi ^2 S_{-1}(z)+\log
   ^2(2) S_{-1}(z)
   \nonumber
    \\
 && 
 \hspace{1.9cm}
 -\log (2) S_2(z)+S_{-2,-1}(z)-\log (2) S_{-1,-1}(-1-z) 
   \nonumber
    \\
 && 
 \hspace{1.9cm}+\log (2)
   S_{-1,-1}(z)-\log (2) S_{-1,1}(-1-z)
   -\log (2) S_{-1,1}(z)
   \nonumber
    \\
 && 
 \hspace{1.9cm}+S_{2,1}(-1-z)-2
   S_{-1,-1,1}(-1-z)-S_{-1,1,-1}(z)
\end{eqnarray}

\begin{eqnarray}
&& S_{-1}(z) S_{1,-1}(-1-z)= \frac{1}{4} \pi ^2 \log (2)-\frac{3 \zeta (3)}{8}+\log (2)
   S_{-2}(-1-z)
   +\log (2) S_{-2}(z)
    \nonumber
    \\
 && 
 \hspace{1.9cm}+\frac{1}{2} \log ^2(2) S_{-1}(-1-z)+\frac{1}{2} \log ^2(2)
   S_{-1}(z)+\frac{1}{12} \pi ^2 S_1(-1-z)
    \nonumber
    \\
 && 
 \hspace{1.9cm}-\frac{1}{2} \log ^2(2) S_1(-1-z)+\frac{1}{12} \pi ^2
   S_1(z)-\frac{3}{2} \log ^2(2) S_1(z)
    \nonumber
    \\
 && 
 \hspace{1.9cm}
 -\log (2) S_2(-1-z)-\log (2)
   S_2(z)+S_{-2,-1}(-1-z)
    \nonumber
    \\
 && 
 \hspace{1.9cm}
 +S_{-2,-1}(z)-2 \log (2) S_{1,-1}(-1-z)-2 \log (2)
   S_{1,-1}(z)
    \nonumber
    \\
 && 
 \hspace{1.9cm}
 -S_{-1,1,-1}(-1-z)-S_{1,-1,-1}(-1-z)-S_{1,-1,-1}(z)
\end{eqnarray}

\begin{eqnarray}
&&
S_{-1}(z) S_{1,1}(-1-z)= -\frac{1}{6} \pi ^2 \log (2)+\frac{\log ^3(2)}{3}+\frac{\zeta
   (3)}{8}+\log (2) S_{-2}(z)
    \nonumber
    \\
 && 
 \hspace{1.9cm}
 -\frac{1}{12} \pi ^2 S_{-1}(-1-z)+\frac{1}{2} \log ^2(2)
   S_{-1}(-1-z)
    \nonumber
    \\
 && 
 \hspace{1.9cm}-\frac{1}{12} \pi ^2 S_{-1}(z)+\frac{1}{2} \log ^2(2) S_{-1}(z)-\frac{1}{12} \pi ^2
   S_1(-1-z)
    \nonumber
    \\
 && 
 \hspace{1.9cm}
 -\frac{1}{2} \log ^2(2) S_1(-1-z)-\frac{1}{2} \log ^2(2) S_1(z)-\log (2)
   S_2(z)
    \nonumber
    \\
 && 
 \hspace{1.9cm}
 +S_{-2,1}(-1-z)-\log (2) S_{1,-1}(-1-z)-\log (2) S_{1,-1}(z)
  \nonumber
    \\
 && 
 \hspace{1.9cm}
 -\log (2) S_{1,1}(-1-z)+\log
   (2) S_{1,1}(z)-S_{2,-1}(z)
    \nonumber
    \\
 && 
 \hspace{1.9cm}
 -S_{-1,1,1}(-1-z)-S_{1,-1,1}(-1-z)+S_{1,1,-1}(z)
\end{eqnarray}

\begin{eqnarray}
&& S_{-1}(z) S_{-1,-1}(-1-z)= -\frac{1}{6} \pi ^2 \log (2)+\frac{4 \log ^3(2)}{3}-\frac{\zeta
   (3)}{4}-\log (2) S_{-2}(-1-z)
   \nonumber
    \\
 && 
 \hspace{1.9cm}
 +\log (2) S_{-2}(z)+\frac{1}{12} \pi ^2 S_{-1}(-1-z)-\frac{1}{12}
   \pi ^2 S_{-1}(z)
      \nonumber
    \\
 && 
 \hspace{1.9cm}
 +2 \log ^2(2) S_{-1}(z)+\log (2) S_2(-1-z)-\log (2) S_2(z)
 \nonumber
    \\
 && 
 \hspace{1.9cm}
 -2 \log (2)
   S_{-1,-1}(-1-z)+2 \log (2) S_{-1,-1}(z)+S_{2,-1}(-1-z)
   \nonumber
    \\
 && 
 \hspace{1.9cm}
 -S_{2,-1}(z)-2
   S_{-1,-1,-1}(-1-z)+S_{-1,-1,-1}(z)
\end{eqnarray}

\begin{eqnarray}
&&  S_1(z) S_{-2}(-1-z) = -\frac{1}{2} \pi ^2 \log (2)+\frac{3 \zeta (3)}{4}-\frac{1}{4} \pi ^2
   S_{-1}(-1-z)
   -\frac{1}{4} \pi ^2 S_{-1}(z)
   \nonumber
    \\
 && 
 \hspace{1.9cm}+\frac{1}{12} \pi ^2 S_1(-1-z)-\frac{1}{12} \pi ^2
   S_1(z)+S_{-2,1}(z)+S_{1,-2}(-1-z)
\end{eqnarray}

\begin{eqnarray}
&& S_1(z) S_2(-1-z)= 3 \zeta (3)-\frac{1}{6} \pi ^2 S_1(-1-z)+\frac{1}{6} \pi ^2
   S_1(z)
    \nonumber
    \\
 && 
 \hspace{1.9cm}
 +S_{1,2}(-1-z)-S_{2,1}(z)
\end{eqnarray}

\begin{eqnarray}
&& S_1(z) S_{-1,1}(-1-z)= -\frac{1}{12} \pi ^2 \log (2)+\frac{2 \log ^3(2)}{3}-\frac{\zeta
   (3)}{8}+\frac{1}{12} \pi ^2 S_{-1}(-1-z)
   \nonumber
    \\
 && 
 \hspace{1.9cm}
 +\frac{1}{2} \log ^2(2) S_{-1}(-1-z)-\frac{1}{4} \pi ^2
   S_{-1}(z)+\frac{1}{2} \log ^2(2) S_{-1}(z)
   \nonumber
    \\
 && 
 \hspace{1.9cm}
 +\frac{1}{12} \pi ^2 S_1(-1-z)-\frac{1}{2} \log ^2(2)
   S_1(-1-z)
   -\frac{1}{12} \pi ^2 S_1(z)
   \nonumber
    \\
 && 
 \hspace{1.9cm}
 +\frac{1}{2} \log ^2(2)
   S_1(z)-S_{-2,1}(-1-z)+S_{-2,1}(z)
   \nonumber
    \\
 && 
 \hspace{1.9cm}
 +S_{-1,1,1}(-1-z)-S_{-1,1,1}(z)+S_{1,-1,1}(-1-z)
\end{eqnarray}

\begin{eqnarray}
&& S_1(z) S_{1,-1}(-1-z)= -\frac{7}{12}  \pi ^2 \log (2)-\frac{\log ^3(2)}{3}+\zeta (3)+\log (2)
   S_{-2}(-1-z)
   \nonumber
    \\
 && 
 \hspace{1.9cm}
 -\frac{1}{12} \pi ^2 S_{-1}(-1-z)-\frac{1}{12} \pi ^2 S_{-1}(z)-\frac{1}{12} \pi ^2
   S_1(z)-\log ^2(2) S_1(z)
   \nonumber
    \\
 && 
 \hspace{1.9cm} 
 -\log (2) S_2(-1-z)+S_{-2,1}(z)-\log (2) S_{1,-1}(-1-z)
 \nonumber
    \\
 && 
 \hspace{1.9cm}
 -\log (2)
   S_{1,-1}(z)+\log (2) S_{1,1}(-1-z)-\log (2) S_{1,1}(z)
   \nonumber
    \\
 && 
 \hspace{1.9cm}
 -S_{2,-1}(-1-z)-S_{1,-1,1}(z)+2
   S_{1,1,-1}(-1-z)
\end{eqnarray}

\begin{eqnarray}
&& S_1(z) S_{1,1}(-1-z)= 3 \zeta (3)+\frac{1}{6} \pi ^2 S_1(-1-z)+\frac{1}{6} \pi ^2
   S_1(z)-S_{2,1}(-1-z)
   \nonumber
    \\
 && 
 \hspace{1.9cm} 
 -S_{2,1}(z)+2 S_{1,1,1}(-1-z)+S_{1,1,1}(z)
\end{eqnarray}

\begin{eqnarray}
&& S_1(z) S_{-1,-1}(-1-z)= \frac{1}{6} \pi ^2 \log (2)+\log ^3(2)+\frac{3 \zeta (3)}{8}-\log (2)
   S_{-2}(-1-z)
   \nonumber
    \\
 && 
 \hspace{1.9cm}
 +\frac{1}{2} \log ^2(2) S_{-1}(-1-z)+\frac{1}{12} \pi ^2 S_{-1}(z)+\frac{3}{2} \log
   ^2(2) S_{-1}(z)
   \nonumber
    \\
 && 
 \hspace{1.9cm}
 -\frac{1}{12} \pi ^2 S_1(-1-z)-\frac{1}{2} \log ^2(2) S_1(-1-z)+\frac{1}{12} \pi
   ^2 S_1(z)
   \nonumber
    \\
 && 
 \hspace{1.9cm}
 +\frac{1}{2} \log ^2(2) S_1(z)+\log (2) S_2(-1-z)-S_{-2,-1}(-1-z)
 \nonumber
    \\
 && 
 \hspace{1.9cm}
 -\log (2)
   S_{-1,-1}(-1-z)+\log (2) S_{-1,-1}(z)+\log (2) S_{-1,1}(-1-z)
   \nonumber
    \\
 && 
 \hspace{1.9cm}
 +\log (2)
   S_{-1,1}(z)-S_{2,1}(z)+S_{-1,-1,1}(z)
   \nonumber
    \\
 && 
 \hspace{1.9cm}
 +S_{-1,1,-1}(-1-z)+S_{1,-1,-1}(-1-z)
\end{eqnarray}

All other other bilinear reflection identities at $w=3$ are obtained by  a change of  the argument $z \leftrightarrow -1-z$.

\subsection{Trilinear reflection identities}\label{sec:trilinear1}

In the previous Section we considered   bilinear reflections obtained by taking a product of two basis harmonic sums at lower weight. 
At weight three one can consider also trilinear reflection identities obtained by taking a product of three basis harmonic sums at weight one listed in eq.~\ref{b1}. There are six of those and all of them can be obtained from a linear combination of the bilinear reflection identities given in Section~\ref{sec:w3} supplemented by the shuffle identities of corresponding harmonic sums.  The trilinear reflection identities are given by

\begin{eqnarray}
&&  S_{-1}(z){}^2 S_{-1}(-1-z) = -\frac{1}{6} \pi ^2 \log (2)+\frac{8 \log ^3(2)}{3}+\frac{\zeta
   (3)}{4}+\log (2) S_{-2}(-1-z)
   \nonumber
    \\
 && 
 \hspace{1.85cm}
 -\log (2) S_{-2}(z)-\frac{1}{12} \pi ^2 S_{-1}(-1-z)+4 \log ^2(2)
   S_{-1}(-1-z)
   \nonumber
    \\
 && 
 \hspace{1.85cm}
 +\frac{1}{4} \pi ^2 S_{-1}(z)-\log (2) S_2(-1-z)+3 \log (2) S_2(z)
 \nonumber
    \\
 && 
 \hspace{1.85cm}
 +4 \log (2)
   S_{-1,-1}(-1-z)
   -4 \log (2) S_{-1,-1}(z)+S_{-1,2}(z)
   \nonumber
    \\
 && 
 \hspace{1.85cm}
 -S_{2,-1}(-1-z)+2 S_{2,-1}(z)+2
   S_{-1,-1,-1}(-1-z)
   \nonumber
    \\
 && 
 \hspace{1.85cm} 
 -4 S_{-1,-1,-1}(z)
\end{eqnarray}

\begin{eqnarray}
&& S_{-1}(z) S_1(z) S_{-1}(-1-z)= \log ^3(2)-\frac{3 \zeta (3)}{8}+\log (2) S_{-2}(-1-z)
 \nonumber
    \\
 && 
 \hspace{1.85cm}
 +2 \log (2)
   S_{-2}(z)+\frac{3}{2} \log ^2(2) S_{-1}(-1-z)-\frac{1}{12} \pi ^2 S_{-1}(z)
    \nonumber
    \\
 && 
 \hspace{1.85cm}
 +\frac{1}{2} \log
   ^2(2) S_{-1}(z)+\frac{1}{12} \pi ^2 S_1(-1-z)-\frac{3}{2} \log ^2(2) S_1(-1-z)
    \nonumber
    \\
 && 
 \hspace{1.85cm}
 +\frac{1}{12} \pi
   ^2 S_1(z)
   -\frac{1}{2} \log ^2(2) S_1(z)-\log (2)
   S_2(-1-z)
    \nonumber
    \\
 && 
 \hspace{1.85cm}
 +S_{-2,-1}(-1-z)+S_{-2,-1}(z)+S_{-1,-2}(z)
  \nonumber
    \\
 && 
 \hspace{1.85cm}
 +\log (2) S_{-1,-1}(-1-z)-\log (2)
   S_{-1,-1}(z)
    \nonumber
    \\
 && 
 \hspace{1.85cm}
 -\log (2) S_{-1,1}(-1-z)-\log (2) S_{-1,1}(z)-2 \log (2) S_{1,-1}(-1-z)
  \nonumber
    \\
 && 
 \hspace{1.85cm}
 -2 \log (2)
   S_{1,-1}(z)+S_{2,1}(z)-2
   S_{-1,-1,1}(z)-S_{-1,1,-1}(-1-z)
    \nonumber
    \\
 && 
 \hspace{1.85cm}
 -S_{-1,1,-1}(z)-S_{1,-1,-1}(-1-z)-S_{1,-1,-1}(z)
\end{eqnarray}

\begin{eqnarray}
  && S_1(z){}^2 S_{-1}(-1-z)= -\frac{1}{6} \pi ^2 \log (2)+\frac{2 \log ^3(2)}{3}+\zeta (3)+\log (2)
   S_{-2}(-1-z)
    \nonumber
    \\
 && 
 \hspace{1.85cm} 
 +\log (2) S_{-2}(z)-\frac{1}{12} \pi ^2 S_{-1}(-1-z)+\log ^2(2)
   S_{-1}(-1-z)
    \nonumber
    \\
 && 
 \hspace{1.85cm} 
 -\frac{1}{12} \pi ^2 S_{-1}(z)+\log ^2(2) S_{-1}(z)-\log ^2(2)
   S_1(-1-z)-\frac{1}{6} \pi ^2 S_1(z)
    \nonumber
    \\
 && 
 \hspace{1.85cm}
 -\log ^2(2) S_1(z)-\log (2) S_2(-1-z)+\log (2) S_2(z)+2
   S_{-2,1}(z)
    \nonumber
    \\
 && 
 \hspace{1.85cm}
 +S_{-1,2}(z)-2 \log (2) S_{1,-1}(-1-z)-2 \log (2) S_{1,-1}(z)
  \nonumber
    \\
 && 
 \hspace{1.85cm}
 +2 \log (2)
   S_{1,1}(-1-z)-2 \log (2) S_{1,1}(z)-S_{2,-1}(-1-z)
    \nonumber
    \\
 && 
 \hspace{1.85cm}
 -2 S_{-1,1,1}(z)-2 S_{1,-1,1}(z)+2
   S_{1,1,-1}(-1-z)
\end{eqnarray}

\begin{eqnarray}
&& S_1(z) S_{-1}(-1-z){}^2 = \frac{1}{3} \pi ^2 \log (2)+2 \log ^3(2)-\frac{9 \zeta (3)}{4}-2 \log
   (2) S_{-2}(-1-z)
   \nonumber
    \\
 && 
 \hspace{1.85cm}
 +\log ^2(2) S_{-1}(-1-z)+\frac{1}{6} \pi ^2 S_{-1}(z)+3 \log ^2(2)
   S_{-1}(z)
   \nonumber
    \\
 && 
 \hspace{1.85cm}
 -\log ^2(2) S_1(-1-z)+\log ^2(2) S_1(z)+2 \log (2) S_2(-1-z)
 \nonumber
    \\
 && 
 \hspace{1.85cm}
 -2 S_{-2,-1}(-1-z)-2 \log
   (2) S_{-1,-1}(-1-z)+2 \log (2) S_{-1,-1}(z)
   \nonumber
    \\
 && 
 \hspace{1.85cm}
 +2 \log (2) S_{-1,1}(-1-z)+2 \log (2)
   S_{-1,1}(z)-S_{1,2}(-1-z)
   \nonumber
    \\
 && 
 \hspace{1.85cm}
 -S_{2,1}(z)+2 S_{-1,-1,1}(z)+2 S_{-1,1,-1}(-1-z)
 \nonumber
    \\
 && 
 \hspace{1.85cm}
 +2 S_{1,-1,-1}(-1-z)
\end{eqnarray}

\begin{eqnarray}
&&   S_1(z){}^2 S_1(-1-z)= 3 \zeta (3)+\frac{1}{6} \pi ^2 S_1(-1-z)+\frac{1}{2} \pi ^2
   S_1(z)-S_{1,2}(z)
    \nonumber
    \\
 && 
 \hspace{1.85cm}
 -S_{2,1}(-1-z)-2 S_{2,1}(z)+2 S_{1,1,1}(-1-z)+4 S_{1,1,1}(z)
\end{eqnarray}

\begin{eqnarray}
&&  S_1(z) S_{-1}(-1-z) S_1(-1-z)= -\frac{1}{6} \pi ^2 \log (2)+\frac{\log ^3(2)}{3}+\frac{\zeta
   (3)}{8}
   \nonumber
    \\
 && 
 \hspace{1.85cm}
 +\log (2) S_{-2}(-1-z)+\frac{1}{4} \pi ^2 S_{-1}(-1-z)+\frac{1}{2} \log ^2(2)
   S_{-1}(-1-z)
   \nonumber
    \\
 && 
 \hspace{1.85cm}
 -\frac{1}{12} \pi ^2 S_{-1}(z)+\frac{1}{2} \log ^2(2) S_{-1}(z)-\frac{1}{2} \log
   ^2(2) S_1(-1-z)
   \nonumber
    \\
 && 
 \hspace{1.85cm}
 -\frac{1}{12} \pi ^2 S_1(z)-\frac{1}{2} \log ^2(2) S_1(z)-\log (2)
   S_2(-1-z)-S_{-2,1}(-1-z)
   \nonumber
    \\
 && 
 \hspace{1.85cm} 
 +S_{-2,1}(z)-S_{1,-2}(-1-z)-\log (2) S_{1,-1}(-1-z)-\log (2)
   S_{1,-1}(z)
   \nonumber
    \\
 && 
 \hspace{1.85cm}
 +\log (2) S_{1,1}(-1-z)-\log (2)
   S_{1,1}(z)-S_{2,-1}(-1-z)
   \nonumber
    \\
 && 
 \hspace{1.85cm}
 +S_{-1,1,1}(-1-z)-S_{-1,1,1}(z)+S_{1,-1,1}(-1-z)
 \nonumber
    \\
 && 
 \hspace{1.85cm}
 -S_{1,-1,1}(z)+2
   S_{1,1,-1}(-1-z)
\end{eqnarray}

All other other trilinear reflection identities at $w=3$ are obtained by  a change of  the argument $z \leftrightarrow -1-z$.

\subsection{The Method}
In deriving the reflection identities presented in this paper we used   Harmonic Sums package by J.~Ablinger~\cite{AblingerThesis}, HPL package by D.~ Maitre~\cite{Maitre:2005uu} and dedicated Mathematica package for pomeron NNLO eigenvalue by N.~ Gromov, F.~Levkovich-Maslyuk and G.~Sizov~\cite{Gromov:2015vua}. 

We expanded around positive and negative integer points  the  product of two harmonic sums $S_{a_1,a_2,..}(z)S_{b_1,b_2,...}(-1-z)$ and the functional basis built of pure Harmonic Sums with constants of relevant weight  listed in Ref.~\cite{AblingerThesis}. Then we compared the coefficients of the irreducible constants of a given weight and solved the resulting set of coefficient equations. We used higher order expansion to cross check our results.   Both the bilinear identities of Section~\ref{sec:w2}, Section~\ref{sec:w3} and the trilinear reflection identities of Section~\ref{sec:trilinear1} are derived from the pole expansion based on the Mellin transform and then checked against the shuffle identities and numerical calculations of the corresponding harmonic sums on the complex plane. 

The reflection identities listed in the present paper are derived for Harmonic Sums analytically continued from even integer values to the complex plane. The analytic continuation from the odd integer values to the complex plane is beyond  the scope of our study.

We attach a Mathematica notebook with our results.

\section{Summary and Discussions}

In this paper we present new reflection identities for the harmonic sums analytically continued to the complex plane. 
The reflection identities at weight two are known for a long time and the reflection identities at weight three present the main result of this paper. 
The need for the identities discussed in this paper emerges in the context of analyzing the general analytic structure of the eigenvalue of the BFKL equation.

\section{Acknowledgements}
We would like to thank Fedor Levkovich-Maslyuk and Mikhail Alfimov for fruitful discussions on details of their calculations in Ref.~\cite{Gromov:2015vua} and Ref.~\cite{Alfimov:2018cms}. We are grateful to Simon Caron-Huot for explaining us the structure of his result in Ref.~\cite{Caron-Huot:2016tzz}.

We are indebted to Jochen Bartels for his hospitality and enlightening discussions during our stay at University of Hamburg where this project was initiated.

\end{document}